\documentclass[a4paper, 10pt]{article}
\usepackage[utf8]{inputenc}
\usepackage{cite}
\usepackage{color}
 \usepackage{latexsym}
  \usepackage{amssymb}
  \usepackage{graphicx}
   \DeclareGraphicsExtensions{.jpg .png .gif .pdf .pdf}
\usepackage[utf8]{inputenc}
\usepackage[T1]{fontenc}
 \usepackage[english]{babel}
  \usepackage{}
  \usepackage{amstext}
  \usepackage{amsmath}
  \usepackage{times} 
  \usepackage{amsfonts}
  \usepackage{cite}
 \date{}

 \title{Linear stability analysis   of Poiseuille flow in porous medium with small suction and injection}
 \author{L. A. Hinvi\footnote{laurent.hinvi@imsp-uac.org}\hspace{0.08cm} 
 A. V. Monwanou\footnote{movins$2008$@yahoo.fr} \footnote{Author to whom correspondence should be addressed:  movins2008@yahoo.fr; vincent.monwanou@imsp-uac.org}\hspace{0.05cm}  and J. B. Chabi
   Orou}
\begin{document}
 \maketitle{Institut de Math\'ematiques et de Sciences
 Physiques,  BP:  613 Porto Novo,  B\'enin}

\begin{abstract}

 We investigate  the  effect of small suction  
 Reynolds number  and permeability  parameter  on the   stability 
 of Poiseuille  fluid flow  in a porous medium between two parallel horizontal 
 stationary porous  plates . We have shown that the  perturbed flow is 
 governed by an equation named modified  Orr-Sommerfeld equation. 
 We find also that the  normalization  of the wall-normal velocity with characteristic 
 small suction (or small injection) velocity is important for a perfect command
 of porous medium fluid flow stability analysis.  The stabilizing effect of the parameters in general  and   
 small suction  Reynolds number  and permeability  parameters  in particular 
    on the linear stability are  found.
\end{abstract}

\section{Introduction}

The flows through porous medium are very much prevalent in nature and therefore, the study of such flows has
become of principal interest in many scientific and engineering applications 
\cite{1, 2, 3, 4, 5, 6, 7, 8, 9, 10, 11, 12, 13}.
This type of flows has shown their
great importance in petroleum engineering to study the movements of natural gas, oil and water through the oil
reservoirs; in chemical engineering for the filtration and water purification processes. Further, to study the
underground water resources and seepage of water in river beds one need the knowledge of the fluid flow through
porous medium. Therefore, there are number of practical uses of the fluid flow through porous media. The porous
medium is in fact a non-homogeneous medium but for the sake of analysis, it may be possible to replace it with a
homogeneous fluid which has dynamical properties equal to those of non-homogeneous continuum. Thus one can
study the flow of hypothetical fluid under the action of the properly averaged external flow and the complicated
problem of the flow through a porous medium reduces to the flow problem of homogeneous fluid with some
additional resistance\cite{1}. In this work, we considered  incompressible viscous fluid flow in porous
medium between two porous plates. We assumed that the plates   are  parallel, horizontal  and stationary.  
 We applied injection at the lower plate and suction at upper see (Hinvi et al.\cite{2}). 
 \par 
 The objective  is to study  the effect 
of  Reynolds number $R_{e\omega}$ and  permeability parameter $Kp$ introduced in 
Navier-Stokes equations  by  the small  suction velocity  and permeability of medium,   on  the stability of 
the Poiseuille  flow in porous medium  as in (Hinvi et al.(\cite{2}and \cite{4}) ).
For  this as in previous paper,  we derive  fourth-order equations named modified fourth-order
Orr-Sommerfeld equations governing the stability analysis of Poiseuille flow  in porous medium.
This allowed us to see  the  simultanous  effect of  $ V^{*}_{\omega}$ or $R_{e\omega}$ and  medium permeability parameter $Kp$
on the stability in the boundary layer.  Thus,  we solve numerically the corresponding eigenvalues
problems. We employ Matlab in all our numerical computations to find eigenvalues.
 Such attempt has been made earlier by  Monwanou et al. \cite{5} the case of the  non-porous 
 plate boundary layers (a flat plate-law)  without
 wall suction by normalizing all the components of the velocity  
with the free stream velocity $U^{*}$.\par
The paper is organized as follows.   
In the second section the mathematical formulation of the problem is made. 
In the third section  the  modified Orr-Sommerfeld equation governing the 
stability analysis  of Poiseuille flow  in   porous medium  is deduced. In the  fourth
section,  the effects of  small injection/suction  parameter $R_{e\omega}$,  
 wave number $k$ and permeability parameter $K_{p}$   on linear  stability  of 
   Poiseuille flow in  porous medium  will be investigated with the help of figures.
 The conclusions will be presented in the
final section.

\section{Mathematical formulation of the problem}
\begin{figure}[htbp]
 \begin{center}
 \includegraphics[width=12cm,   height=6cm]{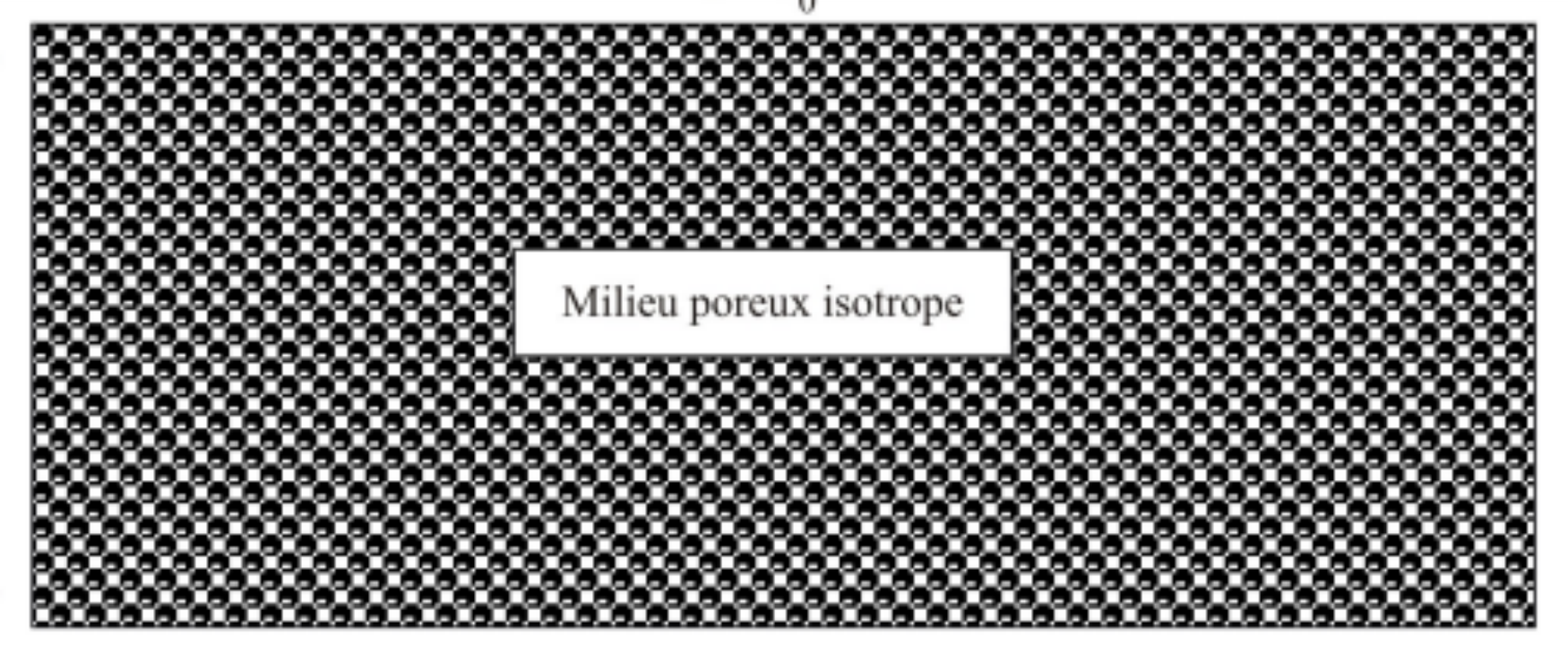}
 \end{center}
 \caption{Porous medium.}
 \label{fig:1}
 \end{figure}

\begin{figure}[htbp]
 \begin{center}
 \includegraphics[width=12cm,   height=8cm]{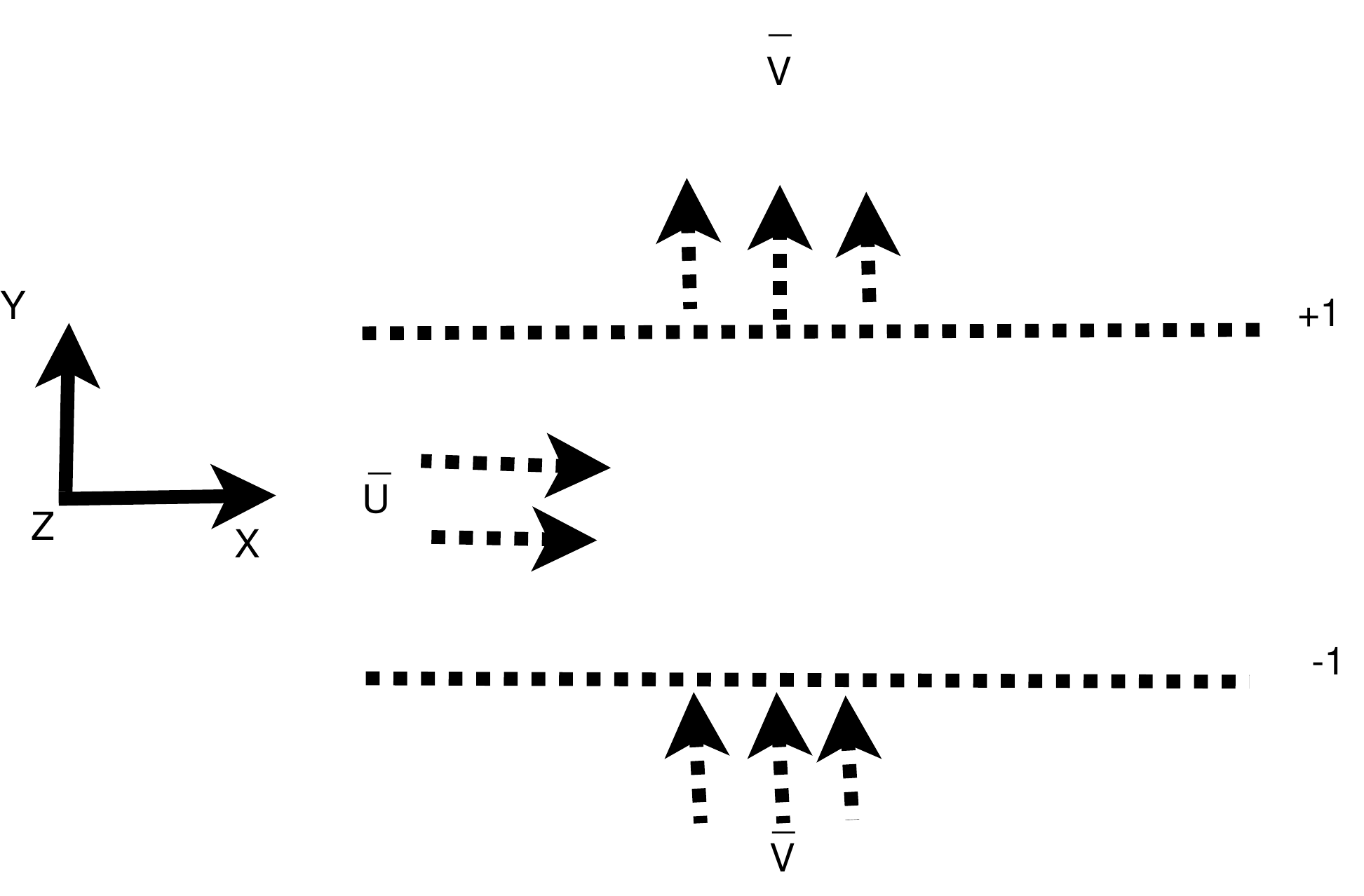} 
 \end{center}
 \caption{Poiseuille porous parallel flat plates flow.} 
 \label{fig:2}
 \end{figure}

We considered a  Poiseuille viscous incompressible fluid flow between  a porous me-dium. We assumed 
 the  medium isotrope see figure \ref{fig:1}. The
 two porous parallel plates  delimiters are assumed  of infinite lengh,   distant $h$. The $y^{*}$ axis  is normal to 
 the planes  of the plates. We considered the simple case where,  the permeability  parameter $Kp$  and 
 velocities suction and injection are constants.
  We work at constant temperature,   the heat transfer aspect of the 
 flow is not studied.  The   small constant 
 injection $V_{\omega}$ is applied at the lower plate and a same small  constant suction $V_{\omega}$,   
 at the upper plate. The two  plates are fixed. We choose the origine on the plane $(x^{*},  0,   z^{*})$ 
 such as $-{h}\leq y^{*}\leq {h}$ and $x^{*}$ parallel to the direction of 
 the    mean fluid flow.  Initially,   $t^{*}<0$,    the fluid
 is assumed to be at rest. When $t^{*}>0$,   the  fluid  starts moving. 
 The porous medium is infact a non-homogeneous medium but for the sake of analysis, it may be possible to
 replace it with a homogeneous continuum. Thus, one can study the flow of hypothetical homogeneous fluid 
 under the action of the properly averaged external forces.  Thus , a complicated problem of the flow through 
 a porous medium reduces to the flow problem of a homogeneous fluid with some resistance.
  Then the equations of fluid flow  in porous medium are simply the equations  of fluid  flow in homogeneous medium
  see (\ref{fig:2})
 added the permeability term. 
 Extra force terms as compared to non-porous  medium fluid are added
on $RHS$ of  Navie-Stokes equations  for homogeneous medium,  due to the permeability medium. 
We can treat porous medium fluid  as
 a single fluid under  assumption that there is a permeability  term to added.\par

The equations of continuity,   motion for the viscous incompressible  fluid in porous medium are
  
\begin{eqnarray}
 \frac{\partial{\tilde{u}_{i}^{*}}}{\partial{\tilde{x}_{i}^{*}}}&=&0, \label{1}\\
 \frac{\partial{\tilde{u}_{i}^{*}}}{\partial{\tilde{t}^{*}}} +
 \tilde{u}_{j}^{*}\frac{\partial{\tilde{u}_{i}^{*}}}{\partial{\tilde{x}_{j}^{*}}} &=&
 -\frac{1}{\rho}\frac{\partial{\tilde{p}}^{*}}{\partial{\tilde{x}_{i}^{*}}}+
 {\nu}\frac{\partial^{2}{\tilde{u}_{i}^{*}}}{{\partial{\tilde{x}_{j}^{*}}}^{2}}-
 \frac{\mu \tilde{u}_{i}^{*} }{\rho k^{*}}.\label{2} 
 \end{eqnarray} 
 
  We introduced the following non-dimensional quantities  as in (Hinvi et al.(\cite{2})  
 
  $$\tilde{x}=\frac{x^{*}}{h}, \qquad  \tilde{y}=\frac{y^{*}}{h}, \qquad  \tilde{z}=\frac{z^{*}}{h},$$

$$ \tilde{u}=\frac{u^{*}}{U},  \qquad
 \tilde{v}=\frac{v^{*}}{V_{\omega}}, \qquad \tilde{w} =\frac{w^{*}}{U},$$

 $$\tilde{t}=\frac{Ut^{*}}{h}, \qquad  \tilde{w} =\frac{w^{*}}{U},  \qquad \tilde{p}=\frac{p^{*}}{\rho U^{2}},$$
 
 $$R_{e}=\frac{U h}{\nu}, \qquad R_{e\omega}=\frac{V_{\omega} h}{\nu}, \qquad K_{p}=\frac{k^{*}}{h^{2}}$$

  Then the Eqs.(\ref{1}) and (\ref{2}) become in cartesien cordinates: 
 \begin{eqnarray}
 \frac{\partial{\tilde{u}}}{\partial{\tilde{x}}}+
 \frac{R_{e\omega}}{R_{e}}\frac{\partial{\tilde{v}}}{\partial{{\tilde{y}}}}+
 \frac{\partial{\tilde{w}}}{\partial{\tilde{z}}}&=&0 \label{3}
 \end{eqnarray}
 \begin{eqnarray}
  \frac{\partial{\tilde{u}}}{\partial{\tilde{t}}}+
  \tilde{u} \frac{\partial{\tilde{u}}}{\partial{\tilde{x}}}+
  \frac{R_{e\omega}}{R_{e}} \tilde{v}\frac{\partial{\tilde{u}}}{\partial{\tilde{y}}}+
  \tilde{w}\frac{\partial{\tilde{u}}}{\partial{\tilde{z}}} &=&
  -\frac{\partial{\tilde{p}}}{\partial{\tilde{x}}}+
  \frac{{\triangledown}^{2}\tilde{u}}{R_{e}}
  -\frac{\tilde{u}}{K_{p}R_{e}} \label{4},  
  \end{eqnarray}
  \begin{eqnarray}
    \frac{\partial{\tilde{v}}}{\partial{\tilde{t}}}+
    \tilde{u} \frac{\partial{\tilde{v}}}{\partial{\tilde{x}}}
 +\frac{R_{e\omega}}{R_{e}} \tilde{v}\frac{\partial{\tilde{v}}}{\partial{\tilde{y}}}+
  \tilde{w}\frac{\partial{\tilde{v}}}{\partial{\tilde{z}}} &=&
  -\frac{R_{e}}{R_{e\omega}}\frac{\partial{\tilde{p}}}{\partial{\tilde{y}}}+
  \frac{{\triangledown}^{2}\tilde{v}}{R_{e}}-\frac{\tilde{v}}{K_{p}R_{e}} \label{5}, 
  \end{eqnarray}
  \begin{eqnarray}
   \frac{\partial{\tilde{w}}}{\partial{\tilde{t}}}+
   \tilde{u} \frac{\partial{\tilde{w}}}{\partial{\tilde{x}}}+
  \frac{R_{e\omega}}{R_{e}} \tilde{v}\frac{\partial{\tilde{w}}}{\partial{\tilde{y}}}+
  \tilde{w}\frac{\partial{\tilde{w}}}{\partial{\tilde{z}}} &=&
  -\frac{\partial{\tilde{p}}}{\partial{\tilde{z}}}+
  \frac{{\triangledown}^{2}\tilde{w}}{R_{e}}-
  \frac{\tilde{w}}{K_{p}R_{e}} \label{6}.
 \end{eqnarray}

\section{The linear disturbance equations}
 The flow is decomposed into the mean flow and the
deviation from it (the disturbance) according to 
\begin{eqnarray}
 \tilde{u}_{i}(r, t)&=& U_{i}(r)+u_{i}(r, t), \label{a1}
 \end{eqnarray}
 \begin{eqnarray}
 \tilde{p}(r, t)&=&P(r)+ p(r, t). \label{a2}
\end{eqnarray}

where $r = (x,  y,  z)$ is the spatial coordinate vector.\par
Inserting (\ref{a1}) and (\ref{a2}) into
the Navier-Stokes equations (\ref{3})- (\ref{6}) we obtain the equations following.

\begin{eqnarray}
  \frac{\partial{u}}{\partial{t}}+U \frac{\partial{u}}{\partial{x}}+u \frac{\partial{U}}{\partial{x}}
  +  \frac{R_{e\omega}V}{R_{e}} \frac{\partial{u}}{\partial{y}}+\frac{R_{e\omega}v}{R_{e}} \frac{\partial{U}}{\partial{y}}+
 {{W}}\frac{\partial{{u}}}{\partial{{z}}}+{{w}}\frac{\partial{{U}}}{\partial{{z}}}
  &=&\nonumber\\
  -\frac{\partial{p}}{\partial{x}}+  \frac{{\triangledown}^{2}u}{R_{e}}-\frac{{u}}{K_{p}R_{e}} \label{7},  
  \end{eqnarray}  

 \begin{eqnarray}
  \frac{\partial{v}}{\partial{t}}+U \frac{\partial{v}}{\partial{x}}+
  u \frac{\partial{V}}{\partial{x}} + \frac{R_{e\omega}V}{R_{e}} \frac{\partial{v}}{\partial{y}}+
  \frac{R_{e\omega}v}{R_{e}} \frac{\partial{V}}{\partial{y}}+
 {W}\frac{\partial{{v}}}{\partial{{z}}}+ {{w}}\frac{\partial{{V}}}{\partial{{z}}}
 &=& \nonumber\\ -\frac{\partial{p}}{\partial{y}}+\frac{{\triangledown}^{2}v}{R_{e}} -\frac{{v}}{K_{p}R_{e}}\label{8},  
  \end{eqnarray}
  
  \begin{eqnarray}
  \frac{\partial{w}}{\partial{t}}+U \frac{\partial{w}}{\partial{x}}+
  u \frac{\partial{W}}{\partial{x}} + \frac{R_{e\omega}V}{R_{e}} \frac{\partial{w}}{\partial{y}} +
  \frac{R_{e\omega}v}{R_{e}} \frac{\partial{W}}{\partial{y}}+
 {{W}}\frac{\partial{{w}}}{\partial{{z}}}+{{w}}\frac{\partial{{W}}}{\partial{{z}}}
  &=&\nonumber\\
  -\frac{\partial{p}}{\partial{z}}+ 
  \frac{{\triangledown}^{2}w}{R_{e}}-\frac{{w}}{K_{p}R_{e}} \label{9},  
  \end{eqnarray}
  
\begin{eqnarray}
  \frac{\partial{{u}}}{\partial{{x}}} +
  \frac{R_{e \omega}}{R_{e}}\frac{\partial{{v}}}{\partial{{y}}}+
 \frac{\partial{{w}}}{\partial{{z}}} &=&0.\label{10}
\end{eqnarray}

 We take  the   dimensional  base flow  for small suction and injection see (Hinvi et al.\cite{2}):
 \begin{eqnarray}
  U(y) &=& (1-y^{2}) \label{a6}\\
  V&=& 1\label{a7}\\
  W&=&0.\label{a8}
 \end{eqnarray}
 To obtain the stability equations for the spatial evolution of three-dimensional,  
 we take the
 dependent on time  disturbances 
\begin{eqnarray}
 (u(x,  y,  z,  t);v(x,  y,  z,  t);  w(x,  y,  z,  t); p(x,  y,  z,  t)); \label{a9}
\end{eqnarray}
 which are  scaled in the same way as above.\par
With the equations (\ref{a6}-\ref{a9}),  the equations (\ref{7}-\ref{10}) become 

 \begin{eqnarray}
  \frac{\partial{u}}{\partial{t}}+U \frac{\partial{u}}{\partial{x}}+
  \frac{R_{e\omega}}{R_{e}} \frac{\partial{u}}{\partial{y}}+
 \frac{R_{e\omega}}{R_{e}}v \frac{\partial{U}}{\partial{y}}
  &=&-\frac{\partial{p}}{\partial{x}}+
  \frac{{\triangledown}^{2}u}{R_{e}}-\frac{{u}}{K_{p}R_{e}} \label{11},  
  \end{eqnarray}
  \begin{eqnarray}
    \frac{\partial{v}}{\partial{t}}+U \frac{\partial{v}}{\partial{x}}+
  \frac{R_{e\omega}}{R_{e}} \frac{\partial{v}}{\partial{y}}
   &=&-\frac{R_{e}}{R_{e\omega}}\frac{\partial{p}}{\partial{y}}+
  \frac{{\triangledown}^{2}v}{R_{e}} -\frac{{v}}{K_{p}R_{e}},  \label{12}
  \end{eqnarray}
  \begin{eqnarray}  
   \frac{\partial{w}}{\partial{t}}+U \frac{\partial{w}}{\partial{x}}+
  \frac{R_{e\omega}}{R_{e}} \frac{\partial{w}}{\partial{y}}
   &=&-\frac{\partial{p}}{\partial{z}}+
  \frac{{\triangledown}^{2}w}{R_{e}}-\frac{{w}}{K_{p}R_{e}} \label{13}.
 \end{eqnarray} 
\begin{eqnarray}
  \frac{\partial{u}}{\partial{x}} +
  \frac{R_{e\omega}}{R_e}\frac{\partial{v}}{\partial{y}}+
 \frac{\partial{w}}{\partial{z}} &=&0. \label{14}
\end{eqnarray}
 The equations (\ref{11}-\ref{14}) above are  the linear disturbance equations 
  that model the  Poi-seuille flow  of the incompressible viscous  fluid in  porous medium
  with  the assumption that there is a constant small  suction at upper plate and a constant small injection at the
lower plate in cartesien coordinates. 
 \section{Modified Orr-Sommerfeld equation}
In this section the modified Orr-Sommerfeld equation governing the 
stability analysis  of Poiseuille flow  in  porous medium is deduced by using Navier-Stokes  equations with the 
same strategy  as ( see Hinvi et al.\cite{2}).
By using continuity equation,  the pressure terms can be eliminated from (\ref{11})-(\ref{14}),  by
reducing the system to two equations for two unknown quantities. For a mean profile (\ref{a6}-\ref{a8}),  
the divergence of (\ref{11})-(\ref{14})
and continuity gives
  \begin{eqnarray}
   {\bigtriangledown}^{2}{p}&=& -2\frac{R_{e\omega}}{R_{e}}\frac{\partial{U}}{\partial{y}}
 \frac{\partial{v}}{\partial{x}}.\label{15}
 \end{eqnarray}

We applied $\bigtriangledown^2$ to the equation (\ref{12}),  unsing (\ref{15})  we found
 \begin{eqnarray}
  \frac{\partial{}}{\partial{t}} {\bigtriangledown}^{2}{v} + 
  {U} \frac{\partial{}}{\partial{x}}{\bigtriangledown}^{2}{v}+
   \frac{R_{e \omega}}{R_{e}}    
   \frac{\partial{}}{\partial{y}}{\bigtriangledown}^{2}{v} -
   \frac{d^{2}U}{{dy}^{2}}\frac{\partial{v}}{\partial{x}}-
   2\frac{dU}{{dy}}\frac{\partial^{2}{v}}{\partial{x}\partial{y}}
 &=& \nonumber\\
 \frac{1}{R_{e}} {\bigtriangledown}^{4}{v}-\frac{\bigtriangledown^2 v}{K_{p}R_{e}}.\label{16}
 \end{eqnarray} 
By neglecting the quadratic derivation (because of linear stability  analysis) the equation (\ref{16}) becomes
 
 \begin{eqnarray}
 \left( \frac{\partial{}}{\partial{t}}+ 
  {U} \frac{\partial{}}{\partial{x}}
   +\frac{1}{K_{p}R_{e}} + \frac{R_{e \omega}}{R_{e}}   
   \frac{\partial{}}{\partial{y}}\right){\bigtriangledown}^{2}{v} -
   \frac{d^{2}U}{{dy}^{2}}\frac{\partial{v}}{\partial{x}}
 &=& \nonumber
 \frac{1}{R_{e}} {\bigtriangledown}^{4}{v}.\label{17}
\end{eqnarray}

 We transform the Eq.(\ref{16}) in an eigenvalue problem.
 Thus, the disturbances are taken 
 to be periodic in time in the streamwise,  spanwise directions,   
 which allow us to assume solutions of the form
 
\begin{eqnarray}
 f(x, y, z, t)&=&\hat{f}(y)e^{i(\alpha x+\beta z-\omega t)}; \label{18}
\end{eqnarray}
where $f$ represents either one of the disturbances $u$,  $v$,  $w$ 
 or $p$ and $\hat{f}$ 
 the amplitude function,  $k$,  $\alpha= k_{x}=k\cos{\theta}$ and  
 $\beta= k_{z}=k\sin{\theta}$ are the wave numbers,  $\omega =\alpha c$ is the
frequency of the wave.
 With  $i^{2}=-1$,   $\theta= (\vec{k_{x}},  \vec{k})$,  $c=c_{r}+ic_{i}$ wave velocity  
 which is taken to be complex,  $\alpha$ and $\beta$ are
 real because of temporal stability analysis considered.

 Then with  the equation (\ref{18}) the equation (\ref{17}) becomes
 \begin{eqnarray}
  i\alpha \left(U-c - i\frac{R_{e \omega}D}{{\alpha}R_{e}}- 
  \frac{i}{{\alpha}K_{p}R_{e}}+i\frac{D^{2} -{k}^{2} }{\alpha R_{e}}
  \right) \left(D^{2}-{k}^2  \right)\hat{v}&=&\nonumber\\ 
 -i\alpha U''\hat{v}    ; \label{19}
 \end{eqnarray}
 where $D=\frac{d}{dy}$.
 
  We have a same boundary conditions   $\hat{v}(\pm 1)=1,\qquad   \hat{v}'(\pm 1)= 0$
as in (Hinvi et al.\cite{2}).\par  
   To transform these boundary conditions to $  \hat{v}_{p}(\pm 1)= \hat{v}_{p}'(\pm 1)$ 
   we make the change, by puting
  \begin{eqnarray}
 v_{p}(x, y, z, t)&=& v(x,y,z,t)-1\label{20}.
\end{eqnarray}
The equation (\ref{19}) becomes then
 \begin{eqnarray}
 \left[\left(U - i\frac{{{R_{e \omega}}}D}{{\alpha} R_{e}}-  \frac{i}{{\alpha}{{K_{p}}}R_{e}}
 +i \frac{D^{2} -{k}^{2} }{\alpha R_{e}} \right) \left(D^{2}-{k}^2  \right) 
 - \left( U''\right)\right]\hat{v}_{p} &=&\nonumber\\
 c \left( D^{2}-k^{2}\right)\hat{v}_{p}\label{21}  
\end{eqnarray}
with boundary conditions for all $(x,  \pm1,   z ,  t>0)$ :
 \begin{equation}
   \hat{v}_{p}(\pm 1)= \hat{v}_{p}'(\pm 1)=0\label{22}.
 \end{equation}

  The  Eq.(\ref{21}) is a flow equation modified by suction Reynolds number 
  ${R_{e \omega}}$ (or the speed of suction and injection) and permeability parameter $K_p$, 
  which  differentiated it, from    the  model study by (Hinvi et al.\cite{2}) by permeability term,
  rewritten as an eigenvalue problem,  where
  $c$ is the eigenvalue and $\hat{v}_{p}$ the eigenfunction.
\begin{equation}
 \left(U - i\frac{{{R_{e \omega}}}D}{{\alpha} R_{e}}-  \frac{i}{{\alpha}{{K_{p}}}R_{e}}
 +i \frac{D^{2} -{k}^{2} }{\alpha R_{e}} \right) \left(D^{2}-{k}^2  \right) 
 - \left( U''\right)\nonumber
\end{equation}
  
 and $  \left( D^{2}-k^{2}\right)$
are the operators.
\section{Stability analysis}
We  consider  three-dimensional disturbances. We use a temporal 
 stability analysis as mentioned above. With $c$ complex as we have defined above, 
 when $c_{i}< 0$ we have stability,  $c_{i}=0$ we have neutral 
 stability and  elsewhere we have  instability.
 We employ Matlab $7.0$  in all our numerical 
 computations to find the eigenvalues.  The mean  flow  profile
  for Poiseuille  flow  in porous medium  is  
 \begin{eqnarray}
 U = \left(1-y^{2},  1,  0 \right) \label{23}
 \end{eqnarray}
 for $R_{e\omega}$ small (i.e. small suction) is considered (see \cite{3})
 The eigenvalue problem (\ref{21})-\ref{22}) is  solved numerically
 with the suitable boundary conditions.
 The solutions are found in a layer bounded at 
 $y=\pm1$.
 The results of calculations are presented in the figures below.
 We present the figures related to the eigenvalue problem.
 For all figure we ploted $c_i=c_i(\mbox{one parameter)}$  by fixing the other parameters.   The 
 yellow color is for $c_{i}=0$.

 \begin{figure}[htbp]
 \begin{center}
 \includegraphics[width=12cm,   height=8cm]{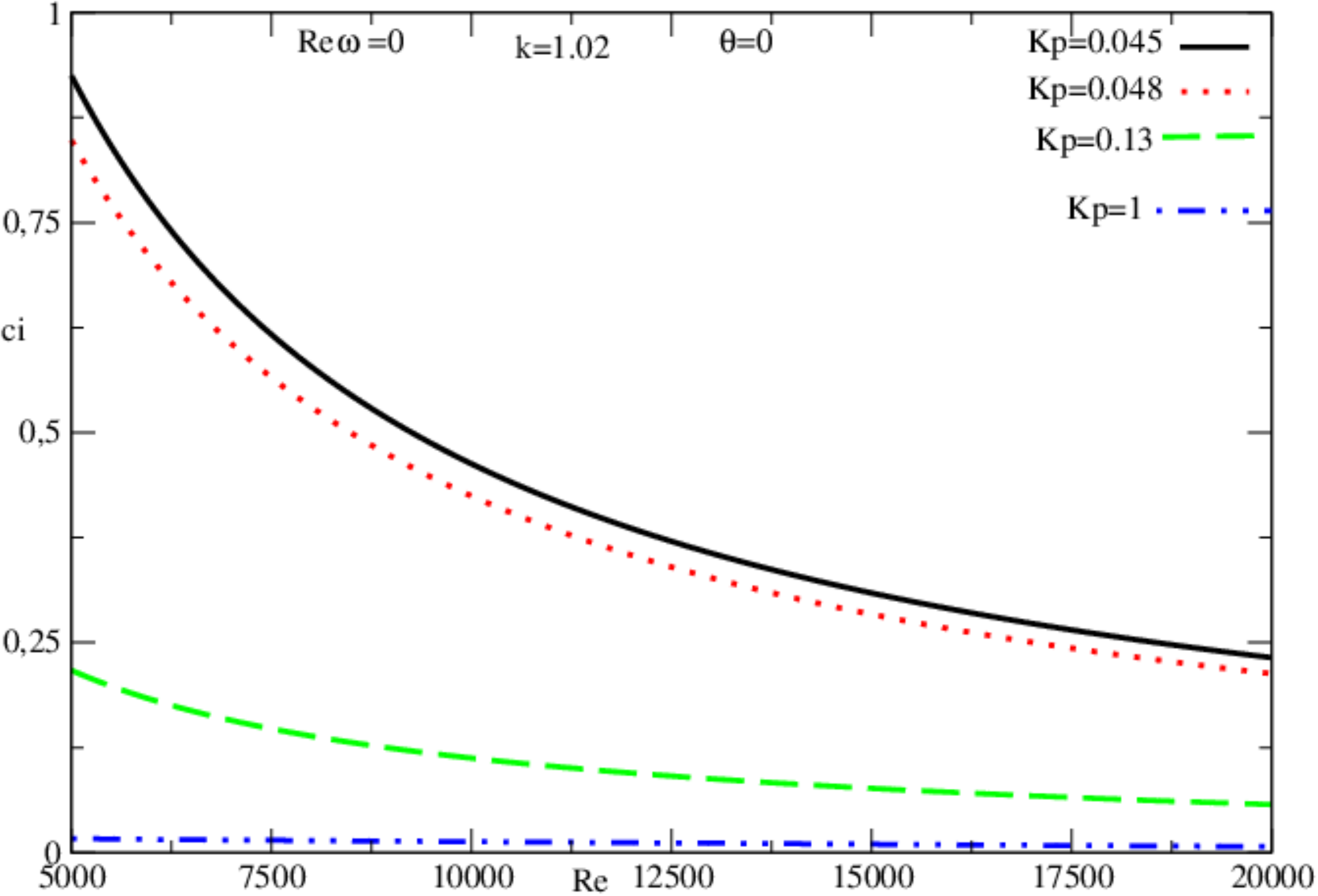}
 \end{center}
\caption{Growth rate $c_{i}$ vs. Reynolds' number for $k=1.02$,  $R_{ew}=0$,  $\theta=0\pi$, 
$Kp=0.045,  0.048,  0.13$ and $1$.}
 \label{fig:s01}
 \end{figure}
  In the figure \ref{fig:s01},  we note for all small  values of permeability parameter that,  $ci>0$. 
  We conclude that the flow is unstable anywhere.
  These curves also tell us that an increase in  the permeability parameter promotes flow stability.
    
 \begin{figure}[htbp]
 \begin{center}
 \includegraphics[width=12cm,   height=8cm]{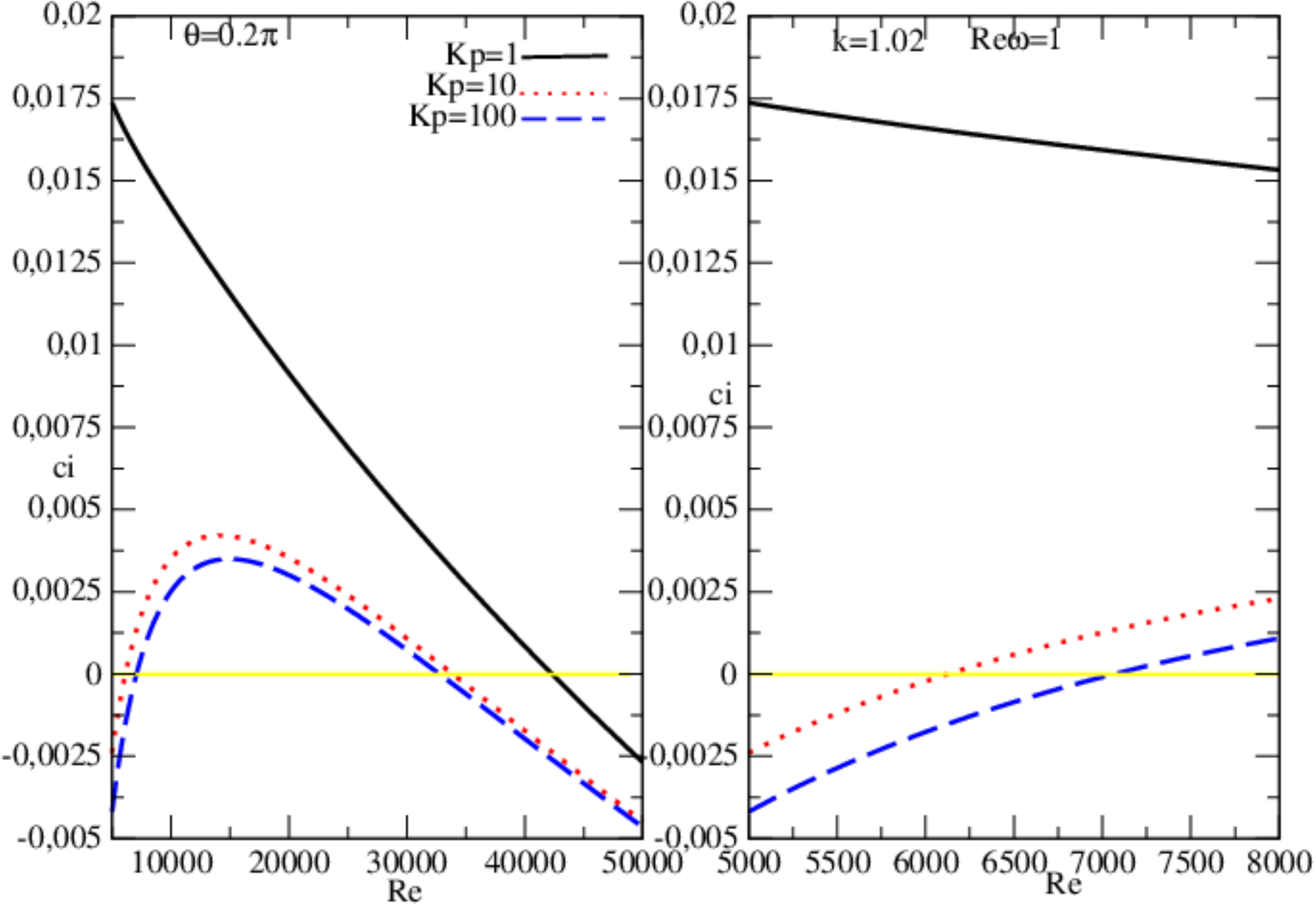}
 \end{center}
\caption{Growth rate $c_{i}$ vs. Reynolds' number $R_e$,  for $k=1.02$,  $R_{ew}=1$,  $\theta=0.2\pi$,  $Kp=1,  10,  100$. 
The right frame is zoom of the left frame.}
 \label{fig:s1zo}
 \end{figure}
 
 Through the curves of the figure \ref{fig:s1zo},   we remark that for  $R_e\leq6200$,   large values of $Kp$ 
 stabilize the flow and for the small value $Kp=1$ the flow is unstable. When $6200 <R_e<33000$  for the all values of $Kp$ 
 the flow is  unstable and  after ($R_e>33000$) it becomes stable elsewhere. The flow stabilizes for small  values  of $ R_e$,  
 becomes unstable and  stabilizes after  for
  $R_e >35000$ i.e we have two transitions of the fluid flow.
  We conclude that  $Kp$'s increasing  stabilizes the    Poiseuille  fluid  flow in porous medium.

 \begin{figure}[htbp]
 \begin{center}
 \includegraphics[width=12cm,   height=8cm]{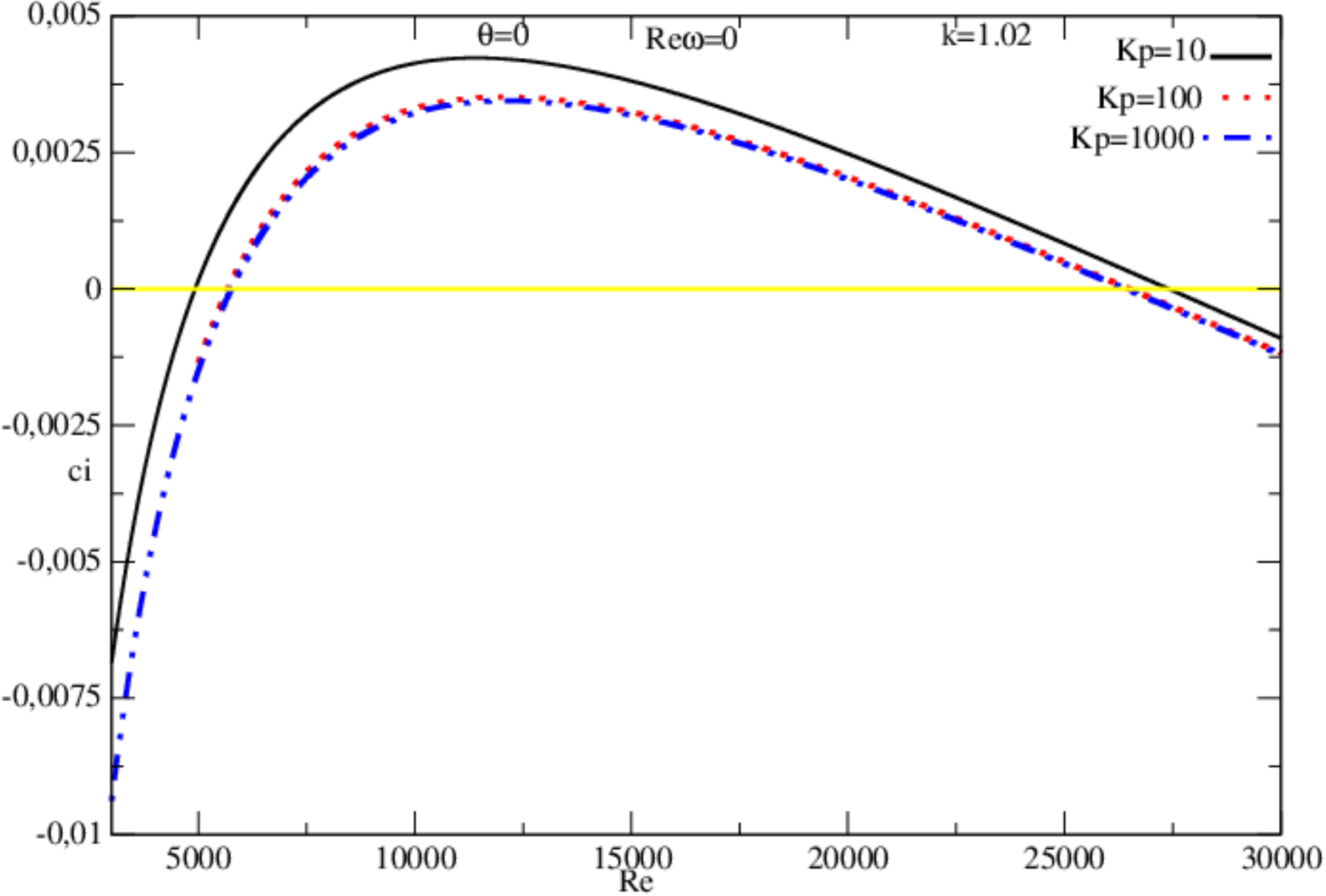}
 \end{center}
\caption{Growth rate $c_{i}$ vs. Reynolds' number for $k=1.02$,  $R_{ew}=0$, 
$\theta=0\pi$,  $Kp=10,  100,  1000$ and $1$.}
 \label{fig:l1}
 \end{figure}
 
  This figure \ref{fig:l1} shows   the behavior of the curve  $c_{i}$ vs. Reynolds' number
  $R_e$ for $k=1.02$,  $R_{ew}=0$,  $\theta=0$,  when $Kp$ increases  
  considerably (strong values of $Kp$). With large values of $Kp$ ($Kp=100,  Kp=1000$) the curves have the same behavior and the 
  limit critical Reynolds' numbers are 
  $  R_{ec}\simeq5770$  and   $ R_{ec}\simeq26500$  respectively for the first and second transitions. 
  In  the equation  (\ref{21}),  we remark that,   when $ Kp\rightarrow \infty$  the permeability term 
  $\frac{1}{\alpha K_p R_e}$ vanishes 
  and the model becomes which has been studed in (Hinvi et al.\cite{2}) where the first critical Reynolds' number is 
  $ R_{ec}\simeq5772.17$. 
  
 \begin{figure}[htbp]
 \begin{center}
 \includegraphics[width=12cm, height=8cm]{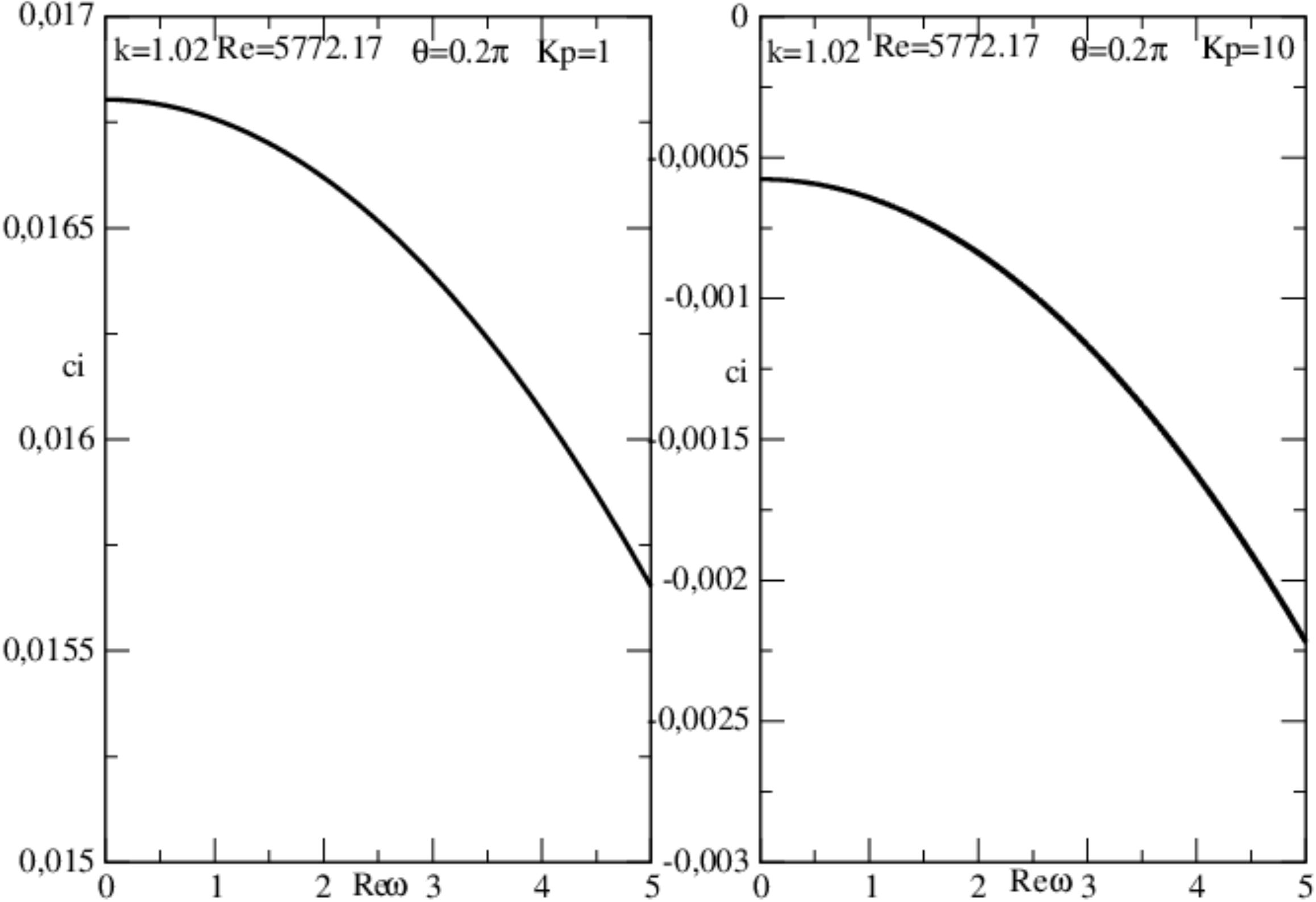}
 \end{center}
\caption{Growth rate $c_{i}$ vs. injection/suction  Reynolds' number $R_{e\omega}$ for
$k=1.02$,  $R_e=5772.17$,  $\theta=0.2\pi$. The right frame ($Kp=1$) and 
 the left frame ($Kp=10$).}
 \label{fig:s2w}
 \end{figure}
  The figure \ref{fig:s2w} show growth rate $c_{i}$ vs. injection/suction  Reynolds' number $R_{e\omega}$ for
$k=1.02$,  $R_e=5772.17$,  $\theta=0.2\pi$,  $Kp=1,  10,  100$.  The both curves are downward,   in  the left frame
 we note $c_i>0$ i.e the flow is unstable. In the right frame  $c_i<0$ then the flow is completely
 stable for this value of $Kp=10$. We conclude  that the suction/injection Reynold's number affects   
 the flow stability and it kept stable for  large permeability parameter value.  
  
 \begin{figure}[htbp]
 \begin{center}
 \includegraphics[width=12cm,   height=8cm]{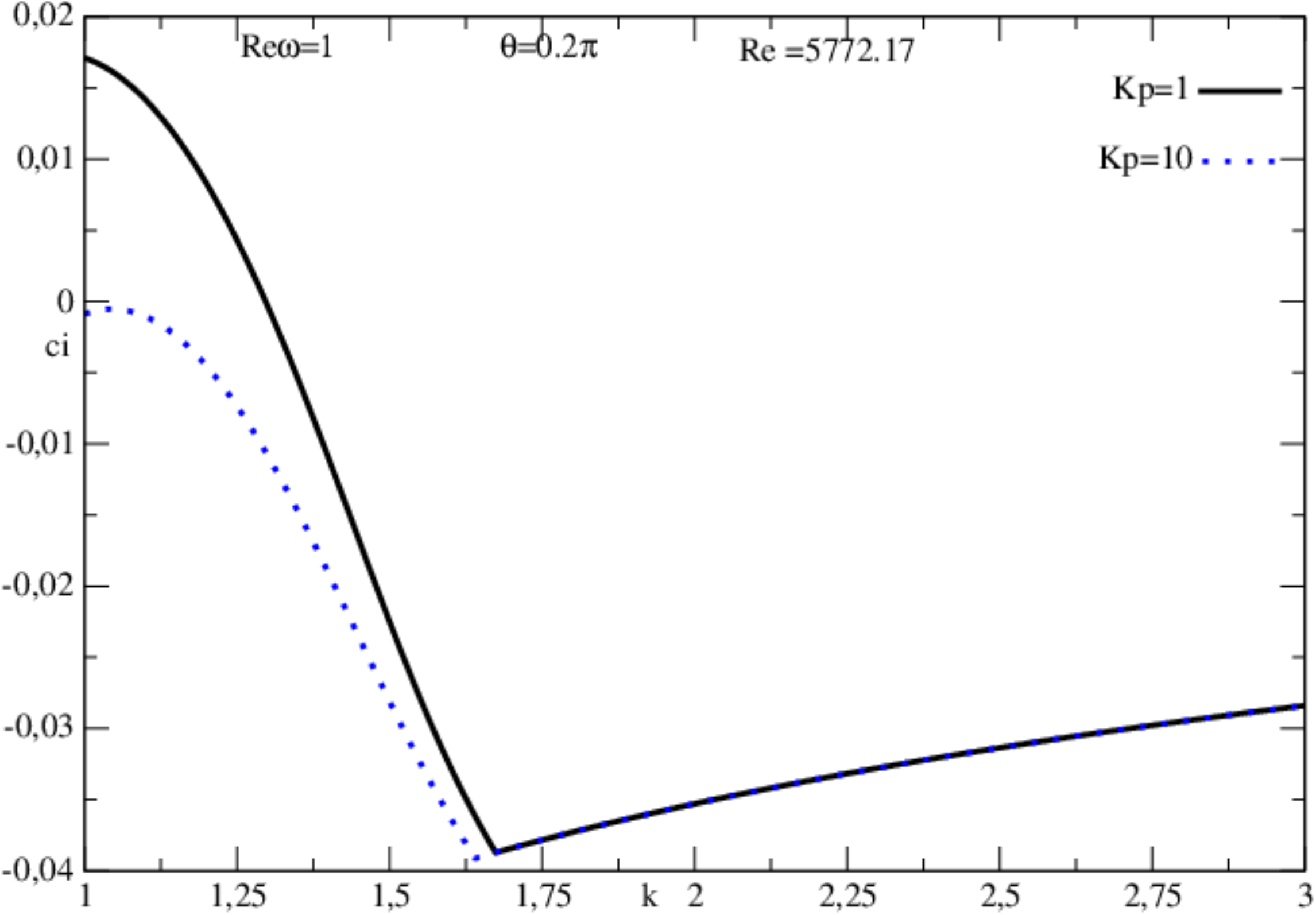}
 \end{center}
\caption{Growth rate $c_{i}$ vs. wave number $k$ for
$Re=5772.17$,  $\theta=0.2\pi$,  $Kp= 1$,  $Kp= 10$ and $R_{e\omega}=1$.}
 \label{fig:s03k}
 \end{figure}
  The curves in figure \ref{fig:s03k} show the appearance of $c_i $
  as a function of wavenumber $k$,  for two different values  of $Kp$. We note that for high values of 
  the permeability parameter,  the flow is completely stable for any order of $k$. 
   Thus for small $Kp$'s values of the order  $1$,  the flow remains unstable
  at the beginning ie for $k <1.5$ and stabilizes after.  We also note that after $ k =  1.73$  the two curves merge.
  We conclude that  there is no special value for $Kp$ to induce any unstability.
 
 \begin{figure}[htbp]
 \begin{center}
 \includegraphics[width=12cm,   height=8cm]{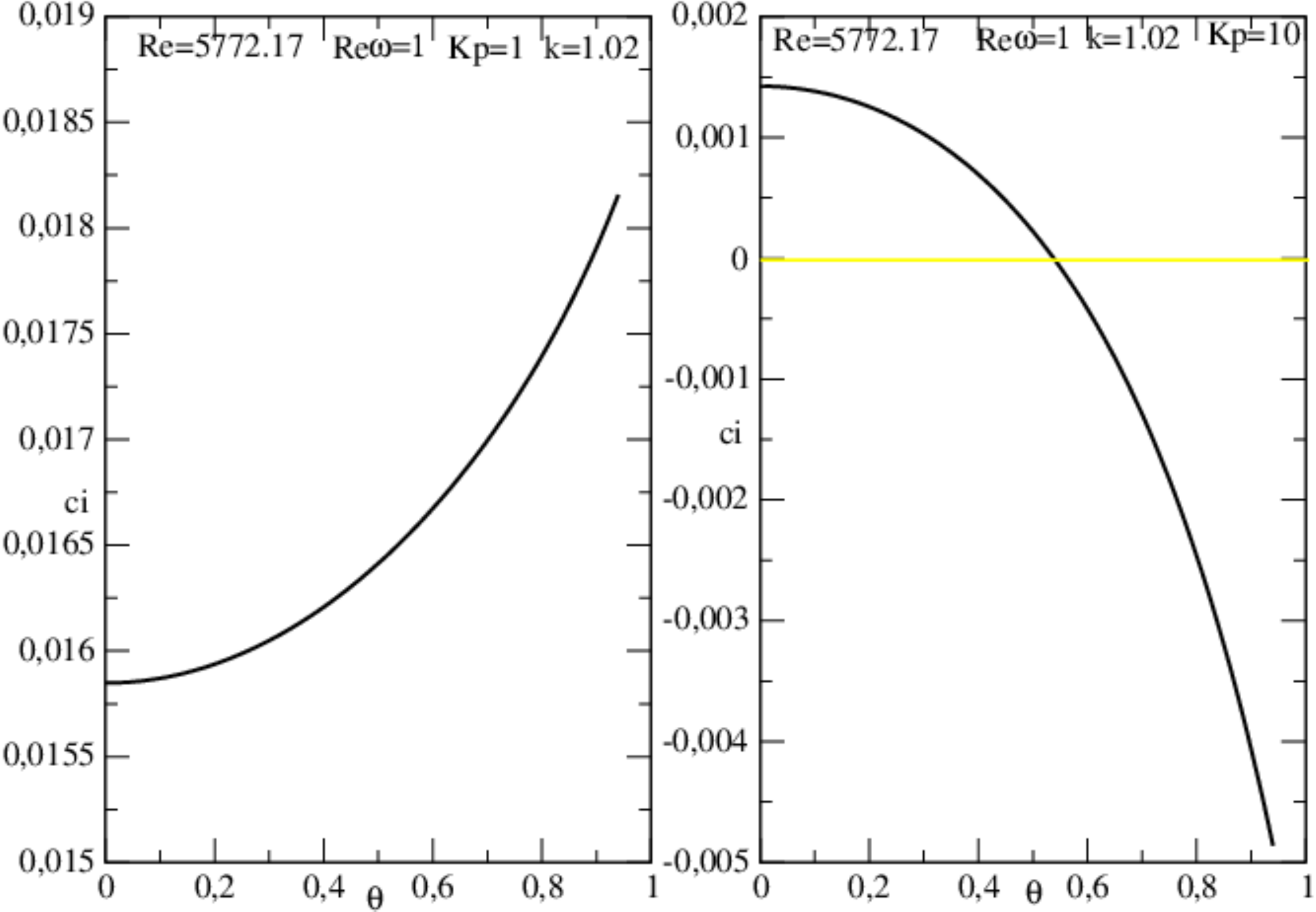}
 \end{center}
\caption{Growth rate $c_{i}$ vs.  $\theta (rad)$ for
$k=1.02$,  $R_e=5772.17$,  $R_{e\omega}=1$. $Kp=1$ for the left frame and $Kp=10$  for the right frame.}
 \label{fig:s4t}
 \end{figure}
  Through  figure \ref{fig:s4t} we have  $c_{i}$ vs. $\theta (rad)$. We note that for $Kp=1$,  $c_i>0$   and
  the curve is ascendent   for all  value of $\theta$ and for $kp=10$  the curve is downward but  we note unstability 
  in beginning and stability at the end.

 \section{Conclusion}
  In this paper,  the study of  the effect 
of  the Reynolds' number $R_{e\omega}$ and  particular the  permeability parameter $Kp$ introduced in 
Orr-Sommerfeld's equation  by  the small  suction velocity  and permeability of medium,   on  the stability of  
the Poiseuille  flow in porous medium has been made. The  modified Orr-Sommerfeld equation 
modeling  Poiseuille flow in porous medium  is deduced. The  growth rates $c_{i}$ vs. the parameters of
system  have been plotted. 
The previous results analysis allowed us to  conclude that all parameters of the system affect
the  Poiseuille flow stability in porous medium and have  a stabilizing effect on the flow.  

\section*{Acknowlegments}
The authors thank IMSP-UAC and Benin gorvernment for financial support.

 \end{document}